\documentclass[%
reprint,prx,
superscriptaddress,
 amsmath,amssymb,
 aps,
floatfix,
]{revtex4-1}
\usepackage{nicefrac}
\usepackage{graphicx,times}% Include figure files
\usepackage{dcolumn}% Align table columns on decimal point
\usepackage{bm}% bold math
\usepackage{hyperref}% add hypertext capabilities
\usepackage{xcolor}
\usepackage[mathlines]{lineno}% Enable numbering of text and display math
\expandafter\ifx\csname package@font\endcsname\relax\else
 \expandafter\expandafter
 \expandafter\usepackage
 \expandafter\expandafter
 \expandafter{\csname package@font\endcsname}%
\fi

\begin{document}
\raggedbottom

%\preprint{APS/123-QED}

\title{Size-regulated symmetry breaking in reaction-diffusion models of developmental transitions}
\author{Jake Cornwall Scoones}
\affiliation{Division of Biology and Biological Engineering, California Institute of Technology, Pasadena, CA 91125, USA}
\altaffiliation{These authors contributed equally to this work}
\author{Deb Sankar Banerjee}
\affiliation{Department of Physics, Carnegie Mellon University, Pittsburgh, PA 15213, USA}
\altaffiliation{These authors contributed equally to this work}
\author{Shiladitya Banerjee}
\thanks{Correspondence: shiladtb@andrew.cmu.edu}
\affiliation{Department of Physics, Carnegie Mellon University, Pittsburgh, PA 15213, USA}


\begin{abstract}
The development of multicellular organisms proceeds through a series of morphogenetic and cell-state transitions, transforming homogeneous zygotes into complex adults by a process of self-organisation. Many of these transitions are achieved by spontaneous symmetry breaking mechanisms, allowing cells and tissues to acquire pattern and polarity by virtue of local interactions without an upstream supply of information. The combined work of theory and experiment has elucidated how these systems break symmetry during developmental transitions. Given such transitions are multiple and their temporal ordering is crucial, an equally important question is how these developmental transitions are coordinated in time. Using a minimal mass-conserved substrate-depletion model for symmetry breaking as our case study, we elucidate mechanisms by which cells and tissues can couple reaction-diffusion driven symmetry breaking to the timing of developmental transitions, arguing that the dependence of patterning mode on system size may be a generic principle by which developing organisms measure time. By analyzing different regimes of our model, simulated on growing domains, we elaborate three distinct behaviours, allowing for clock-, timer-, or switch-like dynamics. By relating these behaviours to experimentally documented case studies of developmental timing, we provide a minimal conceptual framework to interrogate how developing organisms coordinate developmental transitions.
\end{abstract}
\maketitle

\section{Introduction}
In developmental systems, it is important for the mechanistic constituents to `know’ about the size of the living system as a whole \cite{meinhardt2008,green2015}. This is most apparent in developmental transitions, which in many cases only proceed when cells or tissues have reached a critical size. Such control strategies allow living systems to couple developmental time to their size and geometry. What is the physical basis of these phenomena? 

One can envisage two broad classes of size-control mechanisms. The first proposes size-dependent transitions are under external regulation: in developing tissues, this could be manifested as a cell-intrinsic clock, whereby a transition is achieved after a pre-defined time interval \cite{saiz2015}; or a gradient-based mechanism, whereby a distance critically far from the source of signalling molecules triggers a transition \cite{briscoe2015}. Alternatively, size-regulation could be an emergent property of collective decision-making, akin to quorum sensing \cite{waters2005,jorg2019,altschuler2008}: communication between mechanistic constituents allows the system to sense its size. \textit{A priori}, these control mechanisms have several conceptual benefits: synchrony in a transition is more robust, given decisions are made collectively; and such decision making does not rely on a subset of constituents (e.g. source cells in gradient generation), instead being decentralised \cite{chara2014,meinhardt2008}. Can we find examples of emergent size-regulation from collective decision-making in living systems?

Many developmental transitions couple size- and temporal-control to a change in polarity regime: at a critical size, the system may spontaneously break symmetry to polarise, depolarise, bipolarise, or even radically change its pattern. We hypothesise that size regulation of such developmental transitions are an emergent property of the many mechanisms of polarisation. This view provides a framework for understanding developmental time \cite{ebisuya2018}, placing a critical emphasis on system size. 

Here we outline a minimal reaction-diffusion model for size-dependent polarisation in developing systems, arguing that the underlying regulatory motifs can be understood via a substrate-depletion feedback motif coupled with the growth of the system. We then overview cases of size-dependent symmetry breaking across scales, focusing on biochemical systems. We consider size-dependent decision-making within individual cells through to analogous processes in developing multicellular systems, proposing that our minimal model can help unify these divergent processes within a common theoretical framework. We conclude by speculating on the role of these mechanisms in coupling size-dependent transitions to developmental time.

%%%%%%%%%%%%%%%%%%%%%%%%%%%%%%%%%%%%%%%%%%
\section{Reaction-diffusion as a framework to understand size-regulated symmetry-breaking}

Pattern forming reaction-diffusion (RD) systems \cite{turing1952} are widely used to characterise the complex networks of molecular and cellular interactions that underlie biological symmetry breaking. In these systems, pattern formation can arise spontaneously driven by feedback motifs between diffusible molecules (intracellular polarity proteins or extracellular morphogens). Since Turing's insight in 1952, many different RD motifs have been proposed as the physical basis for the emergence of developmental patterns across scales \cite{kondo2010} -- from polarity establishment at the scale of a single cell \cite{butty2002} to pattern formation on the scale of a whole organism \cite{kauffman1981,kondo2010}. 

Perhaps the most famous phrasing of an RD system is the activator-inhibitor circuit, originally developed by Gierer and Meinhardt \cite{meinhardt1974}.  In activator-inhibitor systems, an activator molecule promotes its own production as well as the production of its fast-diffusing inhibitor that suppresses autocatalytic production of the activator. This motif has remarkable explanatory power across contexts, being used to describe the spontaneous establishment of hair follicle spacing \cite{sick2006},  left-right asymmetry establishment in vertebrates  \cite{nakamura2006}, skeletal patterns in growing limbs \cite{newman1979, miura2000,raspopovic2014}, as well as pole-to-pole oscillation of Min proteins during bacterial cell division \cite{meinhardt2001}, and self-organisation of Rho GTPases in the animal cell cortex~\cite{bement2015,michaux2018}. {\it Substrate depletion} models can also yield the spontaneous emergence of periodic patterns~\cite{koch1994,marcon2012}. In these models, the activator consumes its own substrate to promote its autocatalytic production, leading to out-of-phase patterning of the activator and the substrate molecules (Fig.\,\ref{fig:1}a). For example, a substrate-depletion model was been used to explain lung branching, explaining out of phase patterns of gene expression between Shh (the activator) and FGF (the substrate) \cite{menshykau2012}.

Substrate-depletion models are particularly relevant in the study of intracellular pattern formation and polarity establishment as feedback can be phrased in a \textit{mass-conserved} manner. In such models, patterns emerge by the redistribution of polarity proteins~\cite{mori2008, halatek2018}: proteins that form a polarity patch engage in self-recruitment, acting as activators, but this positive-feedback is limited by a finite pool of (typically cytoplasmic) subunits. Variants on mass-conserved substrate depletion models have been used to understand  PAR polarity establishment in the {\it C. elegans} embryo~\cite{munro2004,goehring2011,tostevin2008} and Cdc42 polarisation in {\it S. cerevisiae}~\cite{goryachev2008,savage2012,goryachev2017}. Further, such models exhibit dynamic regimes, for example helping to explain oscillations in the \textit{E. coli} Min-protein system \cite{howard2001,kruse2002,huang2003}.

While the mechanistic constituents and precise feedback architectures of RD mechanisms differ, many rely on a central motif of \textit{local activation and long-range inhibition} \cite{gierer1972}. This concept has acted as an important heuristic in framing models of biological pattern formation, but importantly unifies diverse RD systems within a common mathematical framework. Specifically, recent theoretical models have demonstrated that most RD models of pattern formation can be approximated by the same mathematical formulation: the Swift–Hohenberg equation \cite{hiscock2015}. Strikingly, other mechanisms of periodic pattern formation that rely on cell movement \cite{frohnhofer2013} or mechanical instabilities \cite{murray1984} also rely on local-activation and long-range inhibition \cite{howard2011,hiscock2015}. Hence many dynamical features of these models are applicable across systems and length scales. 

In this perspective, we restrict our focus to RD models for biochemical pattern formation, elucidating the biological significance of a common dynamical feature shared across many motifs: the role of a critical system size for symmetry breaking \cite{koch1994,granero1977,zadorin2017}. To demonstrate this, we develop a mathematical framework for a mass-conserving RD system with a feedback motif similar to activator-substrate models for cell polarisation (Fig.\,\ref{fig:1}a). 

\section{A minimal model for size-regulated symmetry-breaking}

\begin{figure}[htp!]
\includegraphics[width=\columnwidth]{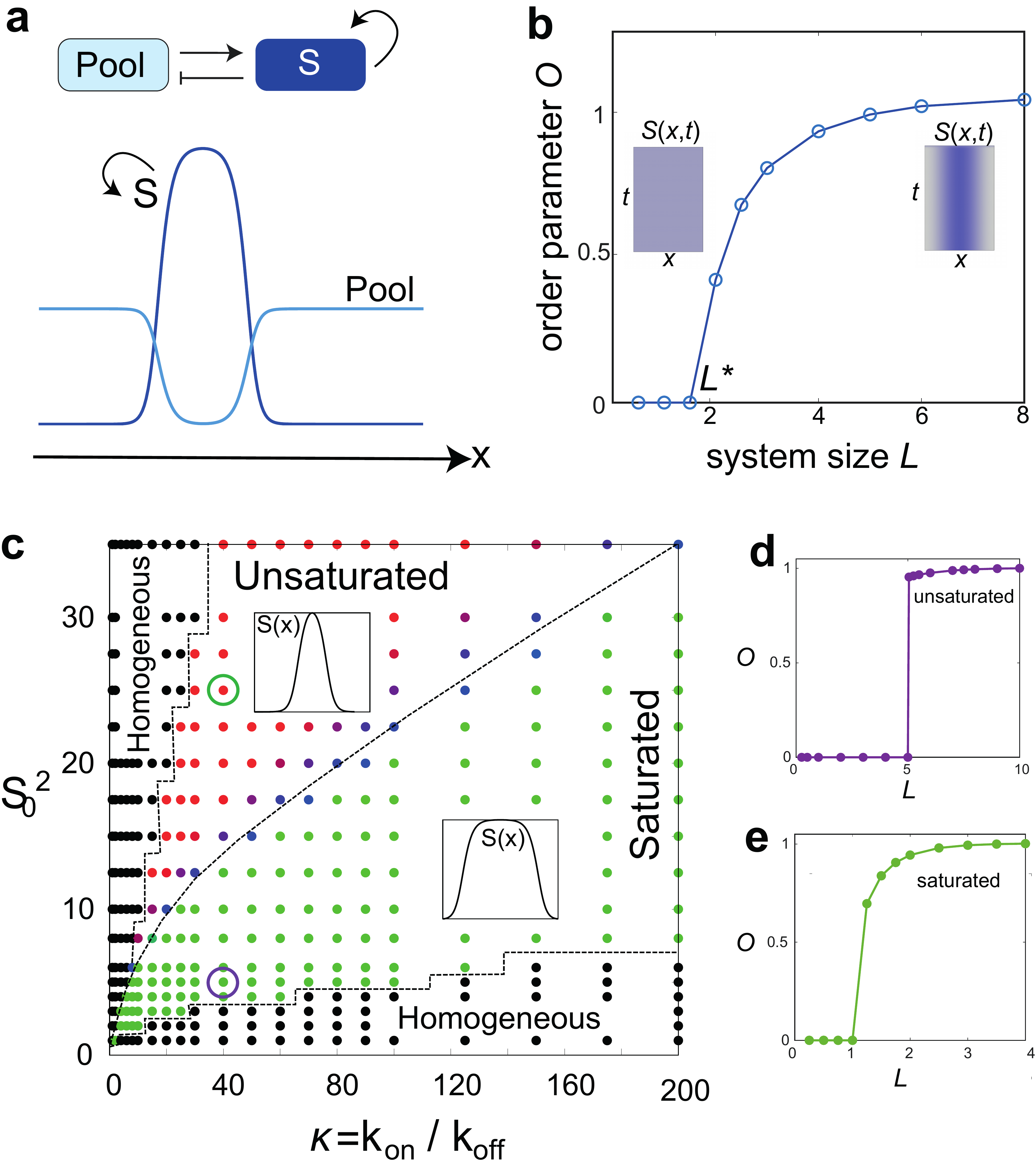}
\caption{{Size regulated symmetry breaking in activator-substrate model. (a) Pattern formation in a model of positive feedback coupled to a finite constituent pool. (b) Patterns form above a critical system size ($L^*$), corresponding to the largest mode where the homogeneous state becomes unstable and the system breaks symmetry. The order parameter $\mathcal{O}$ for symmetry breaking is defined as $\mathcal{O}(L)=\int_0^L \vert dS/dx\vert/\left[\max\limits_{L} \int_0^L \vert dS/dx\vert\right]$, where $\mathcal{O}$ is zero for a homogeneous state and $\mathcal{O}=1$ for a symmetry broken patterned state. All parameters other than system size was kept constant in this analysis and initial perturbations were so chosen that $N/L$ is constant, where $N=\int{P(x,t)+S(x,t)\, dx}$ is the total pool size. Parameters: $D_P=1$, $D_S=0.05$, $k_{\text{on}}/k_{\text{off}}=20$, $S_0^2=10$ and $N/L = 1.5$. (c) Phase diagram in the plane of autocatalytic activity $\kappa$ and the Hill saturation parameter $S_0^2$, showing three different phases: homogeneous state (black), symmetry broken saturated state (green), and symmetry-broken unsaturated state (red). Colormap (green to red) denote the average value of the reaction rate $F_{av}=\int F(x) dx$, computed in the high density region, with $F_{av}=0$ in the saturated state and $F_{av}\neq 0$ in the unsaturated state. $F_{av}=0.01$ (blue points) defines the cross-over value from the saturated state to the unsaturated state. The inset figures show the spatial profiles of $S$ in the different regimes. Parameter values are the same as (b) except for $D_S=0.01$ and $L=3$. (d-e) Order parameter $\mathcal{O}$ for the unsaturated (d) and the saturated (e) regimes, showing that the symmetry of the homogeneous state is broken beyond a critical system size. We used periodic boundary conditions for all numerical simulations, unless otherwise specified. For initial conditions, we assumed a homogeneous $P(x)$ and a sinusoidal $S(x)$ profile of large wavelength.}}
\label{fig:1}
\end{figure}

To analyse the role of system size on the timing of symmetry breaking in a biochemical system, we consider a minimal model for a mass-conserved substrate-depletion system. Specifically we model the spatiotemporal dynamics of a regulatory structure $S$ in a living system of size $L$, and coupled to a finite pool of building blocks. Let $P({\bf x},t)$ denote the concentration of building blocks in the subunit pool at location ${\bf x}$ at time $t$, and $S({\bf x},t)$ is the concentration of building blocks incorporated in the regulatory structure. $S$ increases in amount by depleting the subunit pool $P$, and $S$ can undergo dissociation into $P$ (Fig.\,\ref{fig:1}a). The coupled dynamics of $S$ and $P$ are given by:
\begin{equation}\label{eq:S}
\partial_t S=D_s \nabla^2 S + k_\text{on} P f(S) - k_\text{off} S\;,
\end{equation}
\begin{equation}\label{eq:P}
\partial_t P=D_p \nabla^2 P - k_\text{on} P f(S) + k_\text{off} S\;,
\end{equation}
where $D_p$ and $D_s$ are the diffusion constants of the subunits and the structure $S$ ($D_s\ll D_p$), $k_\text{on}$ parameterises the association rate of subunits to $S$, and $k_\text{off}$ is the constant rate of disassembly of $S$. The function $f$ defines the size-dependence of the autocatalytic production rate of $S$. We assume the functional form $f(S)=S^n/(S_0^n+S^n)$, where the constant $n>1$ controls the strength of cooperative assembly of $S$. The total amount of $P$ and $S$ remains conserved at all times, i.e., $N=\int{P(x,t)+S(x,t)\, dx={\text{constant}}}$, and is assumed to scale linearly with the system size $L$. 

As the structure $S$ grows by locally depleting the pool $P$, localized patterns of $S$ will exist in low density regions of $P$ as long as $D_P \gg D_S$ (Fig.\,\ref{fig:1}a). The symmetry of the homogeneous state breaks and patterns appear above a critical domain size, making this transition size-dependent (Fig. \ref{fig:1}b). The critical size can be obtained from linear stability analysis as
\begin{equation}
 L^* = \left(\frac{2 D_S D_P}{D_S \partial_P{F} + D_P \partial_S{F} }  \right)^{\frac{1}{2}}
\end{equation}
where $F=k_{\text{on}}Pf(S)-k_{\text{off}}S$, and the derivatives are evaluated at the homogeneous steady state. This property of size dependent symmetry-breaking can serve as a decision-making rule to enact developmental state transitions when the system size reaches a critical value $L^*$. As the system size gets larger more discrete structures will emerge (Fig.\,\ref{fig:2}a). Such sequential pattern formation may regulate size-dependent developmental transitions, as we discuss later. 

The $S$-$P$ model introduced above (for $n=2$) is similar to previously studied mass-conserved RD models such as wave-pinning~\cite{mori2008} and Turing-like autocatalytic model~\cite{goryachev2008}. These activator-substrate models exhibit two distinct dynamic regimes~\cite{chiou2018}: the wave-pinning regime, characterised by wide mesa-like patterns and saturated subunit association kinetics; and the Turing regime that yields narrow concentration peaks by virtue of competition between structures. The Turing regime operates below saturation ($S \ll S_0$), where a {\it winner-take-all} competition between structures asymptotically results in a single concentration peak~\cite{otsuji2007,goryachev2008,banerjee2020}. By contrast, in the regime above saturation ($S\gg S_0$), the structures can co-exist for very long timescales, with the timescale of coarsening determined by parameters such as the diffusion coefficients or the reaction fluxes~\cite{brauns2018}. The $S$-$P$ model exhibits both the saturated and the unsaturated regimes that can be obtained by tuning the strength of autocatalytic activity ($\kappa=k_\text{on}/k_\text{off}$) and the Hill saturation parameter $S_0^2$ (Fig.~\ref{fig:1}c). Both these dynamical regimes exhibit size-dependent symmetry breaking (Fig.~\ref{fig:1}d-e), as well as sequential pattern formation for increasing system sizes (Fig.~\ref{fig:2}). In the latter case, we assume that the subunit density is constant for increasing $L$, $\int_0^L (S+P)\,dx \propto L$, as macromolecular contents often scale with cell size~\cite{goehring2012}. For appropriate choice of model parameters in the unsaturated regime, an increase in total subunit pool size coupled with local depletion of the subunit pool gives rise to coexisting peaks of the same height over biologically relevant timescales, much shorter than the long timescale of structure coarsening.

\section{Critical size on polarisation can be utilised to enact cell state transitions }
Our minimal model demonstrates how a positive-feedback motif coupled to features of system size, such as a limiting cytoplasmic pool, can yield size-regulated symmetry breaking of regulatory structures.  In this section we explore biological realisations of our model around the bifurcation from unpolarised to polarised, arguing that cells may utilise these size-dependent properties to coordinate state transitions. 

\subsection{Cell size dependent transition from asymmetric to symmetric division in the early C. elegans embryo}
The polarisation of the early \textit{C. elegans} embryo has become a paradigm in biological symmetry breaking \cite{hoege2013}. Anterior-posterior (AP) polarity in \textit{C. elegans} is established before the first cell division \cite{kemphues1988}. Polarity establishment is achieved by the segregation of two groups of partitioning-defective (PAR) proteins to the anterior versus posterior \cite{munro2004,goehring2011,tostevin2008}. Initially anterior PARs (aPARs) cover the entire membrane of the egg, but upon fertilisation at the posterior, serving as the symmetry breaking cue \cite{goldstein1996}, aPARs segregate anteriorly, and posterior PARs (pPARs) posteriorly. Segregated PARs coordinate polarised division, whereby the division plane is set by the boundary of the two PAR domains \cite{nguyen2007,gonczy2008}. 
%and the segregation of determinants (e.g. P-granules) is coordinated by this polarity \cite{brangwynne2009}. 
\begin{figure}
\includegraphics[width=\columnwidth]{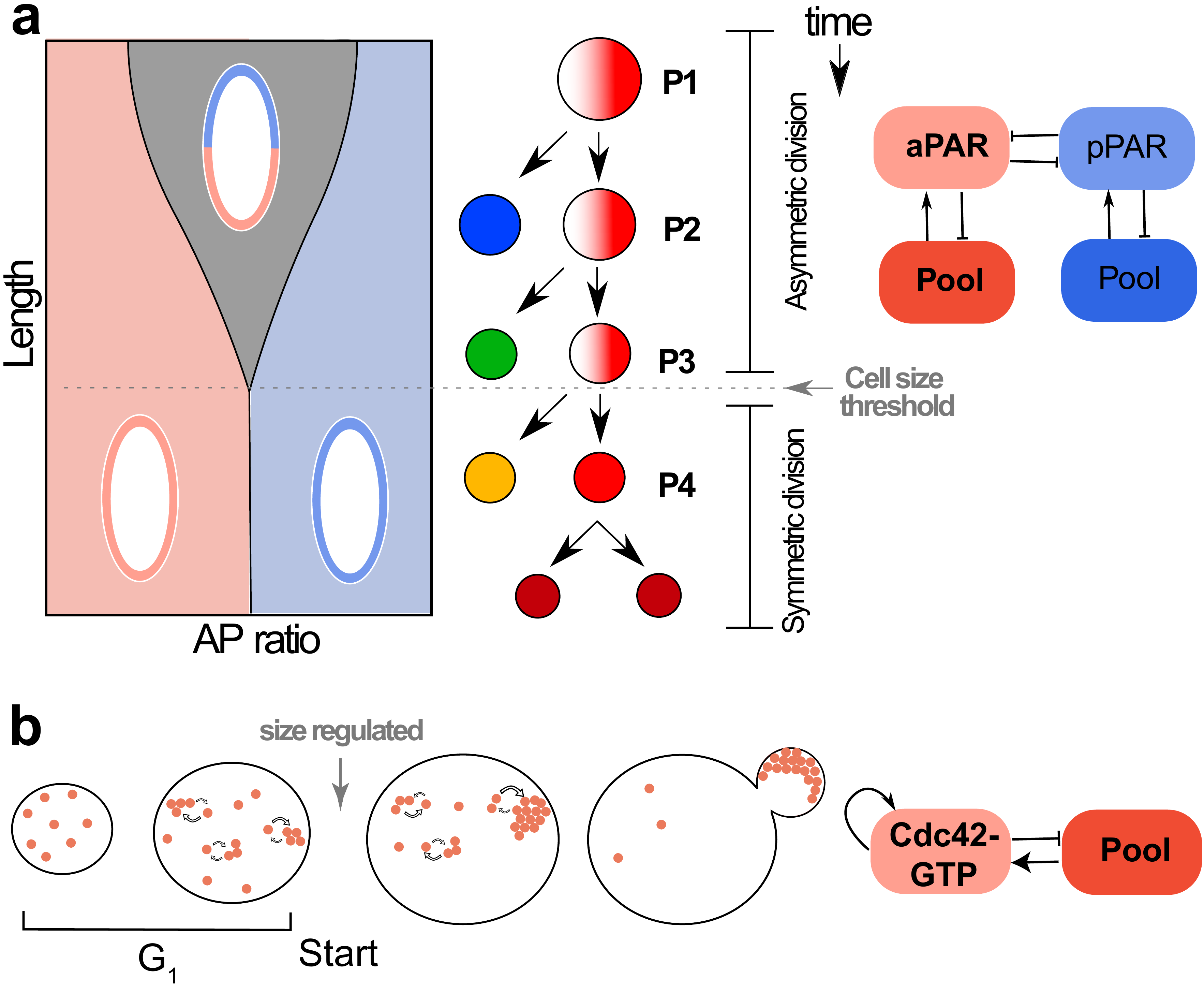}
\caption{{Size-regulated symmetry breaking in single cells. (a) A phase-diagram for the PAR system, considering polarisation state as a function of the circumferential length of the embryo (`Length'), and the ratio of aPAR to pPAR pool size (`AP ratio'). The diagram demonstrates a bipolar state (grey region) becomes unstable below a critical circumferential embryo length (pink and blue regions). Schematics for each of the three states are overlayed, with aPARs  denoted in pink and pPARs  denoted in blue. This bifurcation point quantitatively matches the critical size for which dividing P-cells in the early \textit{C. elegans} embryo transition from asymmetric to symmetric division. Figure adapted from Ref.~\cite{hubatsch2019}. Adjacent is the feedback motif that drives pattern formation. (b) In the budding yeast ({\it S. cerevisiae}), cell cycle commitment to Start is linked to the localisation of Cdc42 effectors at the presumptive bud site~\cite{chiou2017}. Cdc42 polarity establishment is related to the duration of the $G_1$ phase of the cell cycle, which ends at a critical cell size~\cite{di2007}. Models have shown that a growth process with positive feedback leads to Cdc42 polarisation at a single site~\cite{goryachev2008}.  
}}
\label{fig:1b}
\end{figure} 

PARs bind the membrane from a common, finite cytoplasmic pool and diffuse freely. Symmetry breaking is achieved by phosphorylation-dependent mutual inhibition between aPARs and pPARs  \cite{motegi2011,sailer2015}, patterning the cell membrane into two polarisation domains. This double-negative feedback structure reinforces biases in the localisation of PARs and hence plays a similar role as the positive feedback motif in our minimal model. Indeed explicit substrate-depletion models yield phenomenologically identical results \cite{hubatsch2019}. 

The boundary between PAR domains is regulated by the relative diffusivities of aPARs and pPARs, as well as the relative off rates. Boundary length is set by $L_D= \sqrt{D/k_\text{off}}$, where $D$ is the aPAR diffusion constant and $k_\text{off}$ is the aPAR dissociation rate from the membrane. Hence, provided diffusion and dissociation rates are independent of system size, $L_D$ will also be independent of cell size. Modelling confirms that this holds true regardless of structural differences in the model \cite{hubatsch2019}. This length-scale thus sets a minimum cell size that can sustain polarised PAR domains (grey region in Fig.~\ref{fig:1b}a): below this critical size, diffusion overwhelms the capacity for PAR segregation, resulting in homogeneous localisation of either aPARs or pPARs (pink and blue regions of Fig.~\ref{fig:1b}a)

The physical critical size limit may be used by developing \textit{C. elegans} embryos to coordinate a developmental transition. Sequential divisions of generating the first three posterior cells (P1-3) are asymmetric, each generating two daughters of different fates, segregating germline determinants to only one. This pattern shifts to a symmetric mode by the third division of P4 (Fig.~\ref{fig:1b}a), generating the two founding cells of the germline lineage (Z2/Z3) \cite{sulston1983}. Divisions are fast, meaning cell volume declines progressively, falling beneath the theoretical critical cell size for polarisation by P4 \cite{hubatsch2019}. By quantifying division symmetry using 3D reconstructions of PAR distributions, Habatsch et al. \cite{hubatsch2019} found that the polarisation regime shifts from asymmetric to symmetric by P4. The timing of this regime shift can be changed by reducing embryo size through genetic (ima3 RNAi) or physical (laser mediated extrusion) perturbations. Thus a reduction in cell size coordinates a developmental transition in \textit{C. elegans} embryos from asymmetric to symmetric cell division.

\subsection{Size-dependent polarity establishment in budding yeast}
\begin{figure*}
\includegraphics[width=0.8\linewidth]{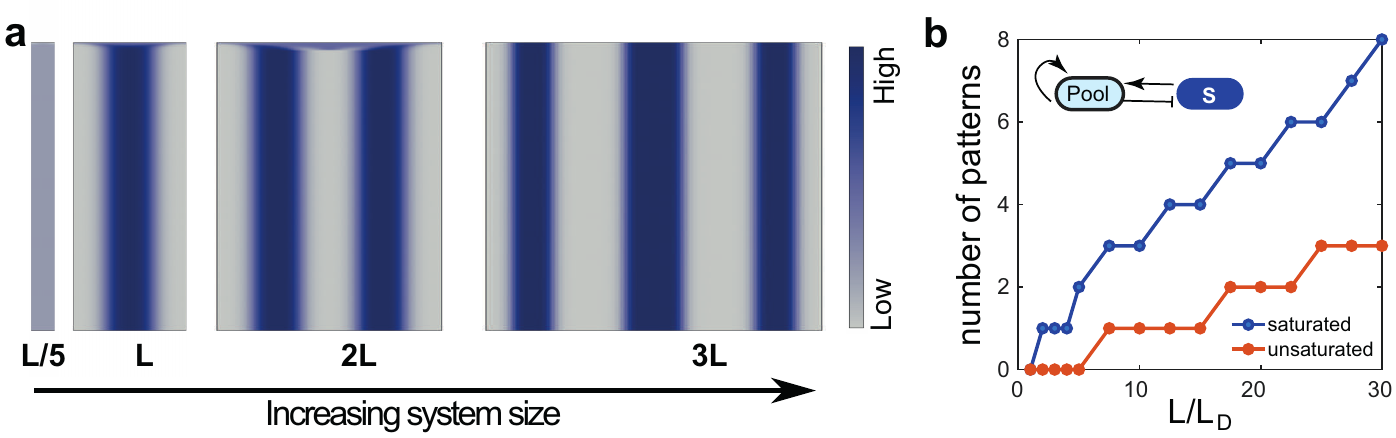}
\caption{{Sequential pattern formation in growing domains.} 
(a) Kymographs of $S(x,t)$ for increasing system size. Figures show spontaneous pattern formation via symmetry breaking of the homogeneous state above a critical system size $L^*$. As we simulate larger systems (Eq.~\eqref{eq:S}-\eqref{eq:P}), multiple patterns emerge in a size dependent manner. Simulation was done as described in Fig.\,\ref{fig:1}b. (b) Number of {\it patterns} as a function of system size, for both the saturated and unsaturated regimes of the $S$-$P$ model. Parameter values for (a) and (b) are the same as Fig.~\ref{fig:1}b and Fig.~\ref{fig:1}c, respectively. Small amplitude random (uniform distribution) initial conditions for $S$ and $P$ were used for (b). }
\label{fig:2}
\end{figure*}

Cell cycle commitment to budding in {\it S. cerevisiae} follows from Cdc42 polarity establishment at the presumptive budding site (Fig.\,\ref{fig:1b}b). The small Rho GTPase Cdc42-GTP forms a polarity patch to mark the bud location \cite{ozbudak2005, okada2013}. This polarity pattern emerges from an autocatalytic positive feedback via Bem1 on the clustering of slowly diffusing membrane bound Cdc42-GTP \cite{butty2002}, while the cytosolic Cdc42-GDP diffuses fast (Fig.\,\ref{fig:1b}b). Polarity establishment in the system can be captured by mass-conserved substrate-depletion model, with a slow-diffusing activator and a fast-diffusing substrate~\cite{otsuji2007,goryachev2008,mori2008}. With appropriate choice of parameters, activator-substrate models would predict the formation of single polarity cluster beyond a critical cell size~\cite{chiou2018} (Fig.\,\ref{fig:1}). Thus the establishment of Cdc42 polarisation can be linked to a critical cell size, consistent with models of critical cell size threshold at the termination of G$_1$ phase of the cell cycle~\cite{di2007,turner2012}.
%This chemical reaction network will give rise to a single polarity pattern above a critical cell size and thus can provide a mechanism of cell size homeostasis as the $G_1$ phase of the cell cycle lasts until the polarity establishment (Fig.\,\ref{fig:1}d) which is regulated by the cell size-dependent symmetry breaking. 
Molecular rewiring experiments have shown that when the Bem1 is tweaked to diffuse very slowly, multiple Cdc42 polarity patches are formed \cite{howell2009}. Since the onset of pattern formation depends on $L/L_D$, with $L_D$ the diffusion length, slowing down diffusion is equivalent to increasing the system size, so multiple patterns emerge in accordance with predictions from increasing domain length in our minimal RD model (Fig.~\ref{fig:2}a-b).

% {\color{red}
% \subsection{Size-dependent polarity establishment in budding yeast}
% }

\section{Sequential pattern formation and polarisation can be coordinated by a growing domain}
Across scales of biological organisation, structures of a bipolar or iterative nature are abundant, and their development is often sequential. While clock-and-wavefront models have successfully explained sequential patterning of axial segmentation in vertebrates \cite{maroto2012} and recently also some invertebrates \cite{sarrazin2012}, sequential pattern formation can also arise from collective decision-making. Our minimal mass-conserved substrate-depletion model illustrates how an RD mechanism can give rise to sequential periodic patterning: smaller domain sizes can sustain only a single pattern via an instability of the homogeneous state, whereas domain growth provides sufficient space to accommodate multiple patterns (Fig. \ref{fig:2}a-b). Here we implicitly assume that the RD system relaxes much faster than the timescale of domain growth, and that subunit concentration remains unchanged. When domain growth rate is comparable to the reaction rate, different dynamic patterns emerge as discussed in Section 6. In spite of differences in their mechanistic bases, sequential patterning through domain growth is common among many RD models (activator-inhibitor and substrate-depletion) \cite{koch1994}. In the following, we expand our focus beyond just mass-conserved models, to illustrate how local-activation and long-range inhibition can explain how growth can couple developmental tempo to state transitions.

\subsection{Neuronal sequential bipolarisation coordinated by membrane growth }
Neuronal polarisation is critical for brain development. Polarisation commences as soon as neurones complete their final division, by a process of neurite formation and selection. \textit{In vitro} studies have suggested neurones acquire a bipolar phenotype, generating a leading neurite, key in guiding migration, and a trailing neurite which later acquires axonal fate \cite{loturco2006,noctor2004}. Bipolarisation \textit{in vitro} is achieved stochastically, whereby the position of the first neurite is seemingly random, with the second being positioned opposite to the first \cite{de2008}. How are these patterns coordinated? 

Menchón et al. \cite{menchon2011} proposed an activator-inhibitor Turing model for cell polarisation. They argued that the necessary feedback architecture for a Turing instability is manifest in developing neurones: integral membrane proteins (the polarisation cue) undergo cooperative self-recruitment i.e. local-activation; and also recruit more diffusive endocytosis modulators which facilitate their removal i.e. long-range inhibition (Fig.~\ref{fig:2b}a). Indeed, in the right parameter regime in a finite domain, simulations suggest neurones can spontaneously break symmetry. The polarity regime is critically dependent on membrane size: a subcritical size prohibits symmetry breaking (like in \textit{C. elegans}); an intermediate size allows for a single polarity axis; and larger sizes allow for a bipolar phenotype (Fig.~\ref{fig:2b}a). Sequential and "mirrored" polarisation can be achieved by membrane growth. A growing domain leads to a time-dependent bifurcation, whereby the cell transitions from a unipolar to bipolar stability regime. The "mirroring" of the second neurite on the first can be rationalised in terms of the feedback circuit: the region of membrane furthest from the first neurite will display the lowest concentration of inhibitor. Neurones thus coordinate the developmental timing of bipolarity through a size-dependent process. %Indeed this ordering may be critical for development, both in terms of the underpinning cell biology — the first neurite is thought to give rise to the future axon \cite{de2005} — and overarching principles of robust self-organisation — sequential patterning modes are thought to generate more regular patterns \cite{koch1994}. 

\subsection{Size dependent sequential patterning in mammalian development -- Insights from gastruloids}
Unlike in \textit{C. elegans}, establishment of anterior-posterior polarity in the epiblast of mammalian embryos occurs well after the first cell division, an axis that lays the ground plan for the commitment of germ-layers during gastrulation. In mice, AP symmetry breaking has long thought to be coordinated by the positioning of extra-embryonic cues to the posterior and hence specifying the future primitive streak \cite{arnold2009}.
%the anterior visceral endoderm (AVE) secretes antagonists of key morphogens including Wnt and Nodal, limiting signalling to the posterior and hence specifying the future primitive streak \cite{arnold2009}. 
This view of sequential polarity hand-off has been thrown into question in recent years by several \textit{in vitro} systems, suggesting that epiblast has the capacity to break symmetry spontaneously in the absence of extra-embryonic cues \cite{harrison2017,van2014,warmflash2014,ten2008,sagy2019}. While the precise genetic constituents of this symmetry breaking are under contention \cite{turner2017,morgani2019}, several that argue some form of reaction-diffusion system is at play, citing for example the co-expression of morphogens with their extra-cellular antagonists, e.g. Wnt and its antagonist Dkk \cite{juan2001,shahbazi2019}. 

\begin{figure}
\includegraphics[width=\columnwidth]{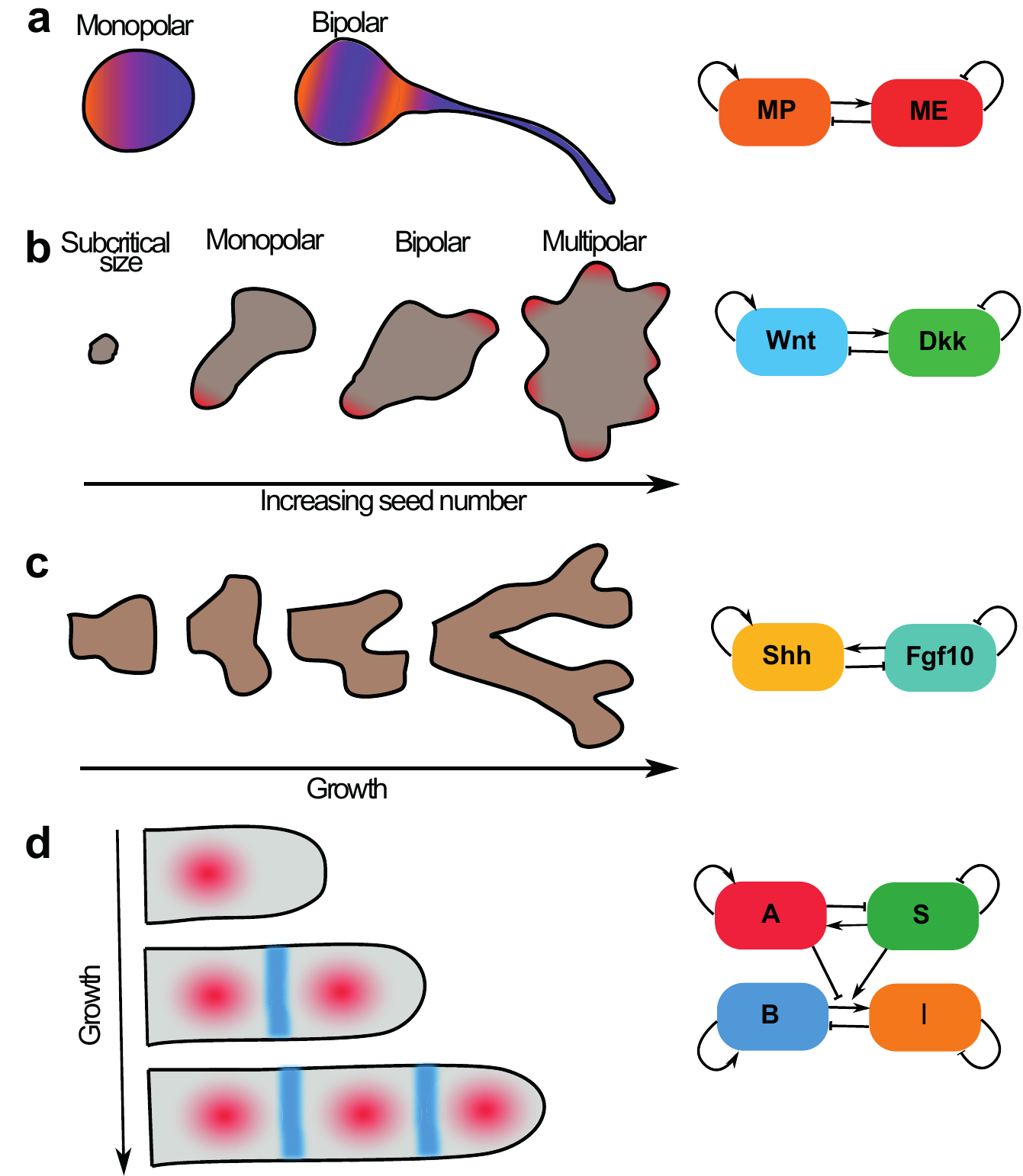}
\caption{{Sequential pattern formation in developmental systems.} (a) In \textit{in vitro} cultured neurones, polarity arises sequentially. A second polarity axis is formed after cell growth, and its orientation is ``mirrored'' off the first. A putative symmetry breaking circuit is presented adjacently \cite{menchon2011}, considering a membrane-protein (MP) activator coupled to modulators of endocytosis (ME), representing the effects of small GTPases. (b) Gastruloids polarise and elongate only when initialised with a critical number of cells \cite{van2014}. For seed numbers beyond this initial bifurcation value, gastruloids can self-organise more axes. T/Brachyury expression is localised to the protrusion in monopolarised gastruloids, and is speculated to also be localised to further protrusions in multipolar variants. A potential feedback circuit is drawn adjacently, which remains to be investigated. (c) An activator-substrate model for lung branching~\cite{menshykau2012}, based on autocatalytic production of the signalling molecule Shh (activator) at the lung bud tip, via consumption of the substrate molecule Fgf10. (d) A dot-stripe mechanism is proposed to pattern the joints of developing digits: a Turing-like dot-forming system specifies the positions of bones and orients through repression a Turing-like stripe-forming system to specify joints. Modelled on a growing domain, sequential joint specification emerges, with joints forming near the developing tip. A coupled-Turing scheme is described adjacent, considering a dot-forming substrate-depletion module (A,S) coupled to a stripe-forming activator-inhibitor module (B,I).}
\label{fig:2b}
\end{figure} 

Consistent with this hypothesis, \textit{in vitro} systems display size-dependence in symmetry breaking capacity and patterning modality \cite{sagy2019,van2014}. This is seen in gastruloids, small aggregates of embryonic stem cells (ESCs) that can spontaneously break symmetry \cite{van2014}, axially elongating and displaying polarised expression of primitive streak marker T/Brachyury. 
%which have also recently been shown to display collinear rostrocaudal Hox expression \cite{beccari2018} and morphological transformations akin to somitogenesis \cite{van2020}. 
In refining their protocol for generating gastruloids, van den Brink et al. \cite{van2014} found that seeding microwells with different numbers of ESCs yielded qualitatively different phenomenology (Fig.~\ref{fig:2b}b): critically small ($\leq 200$ cells) aggregates could not break symmetry; aggregates of intermediate size ($\sim 300-400$ cells) could subsume a unipolar state; aggregates of double that size ($\sim 800$ cells) displayed two oppositely positioned poles; and critically large aggregates ($>1600$ cells) generated many poles. These results are consistent with a Turing-like system controlling polarity, whereby critically small domains cannot sustain an instability, whereas increasingly large domains can maintain progressively more 'peaks’. This is furthered by the observation that bipolar gastruloids polarise sequentially, with the second pole seemingly emerging after growth, protruding from the opposite edge of the structure.

\subsection{Sequential patterning of phalanges in developing digits is coordinated by coupling patterning to growth}
An analogous mechanism may explain the sequential specification of joints in developing digits of tetrapod limbs. Digit patterning and growth are concomitant, with joints being laid down sequentially as progenitor cells are added to the distal tip. Guided by gene expression patterns, as well as mutant phenotypes, joint patterning has been proposed to be governed by a coupled Turing system \cite{cornwallscoones2020} (Fig.~\ref{fig:2b}d). An activator-substrate system specifies the positions of bones (phalanges) by prescribing a series of 'dots' of gene-expression; which represses a second activator-inhibitor system to specify joints as 'stripes' of gene expression at alternate positions. Simulations on a static domain recapitulate both wild-type and mutant expression patterns, but patterning occurs simultaneously across the entire digit. However, simulated on a growing domain, adding new cells distally, leads to a shift in dynamics in favour of sequential patterning. Hence here too, the coordination of developmental timing and growth may be an emergent property of patterning by collective decision-making.

%%%%%%%%%%%%%%%%%%%%%%%%%%%%%%%%%%%%%%%%%%%%%%%%%%%%%%
\section{Regulating pattern size and lifetime in growing systems}

\begin{figure}
\includegraphics[width=\columnwidth]{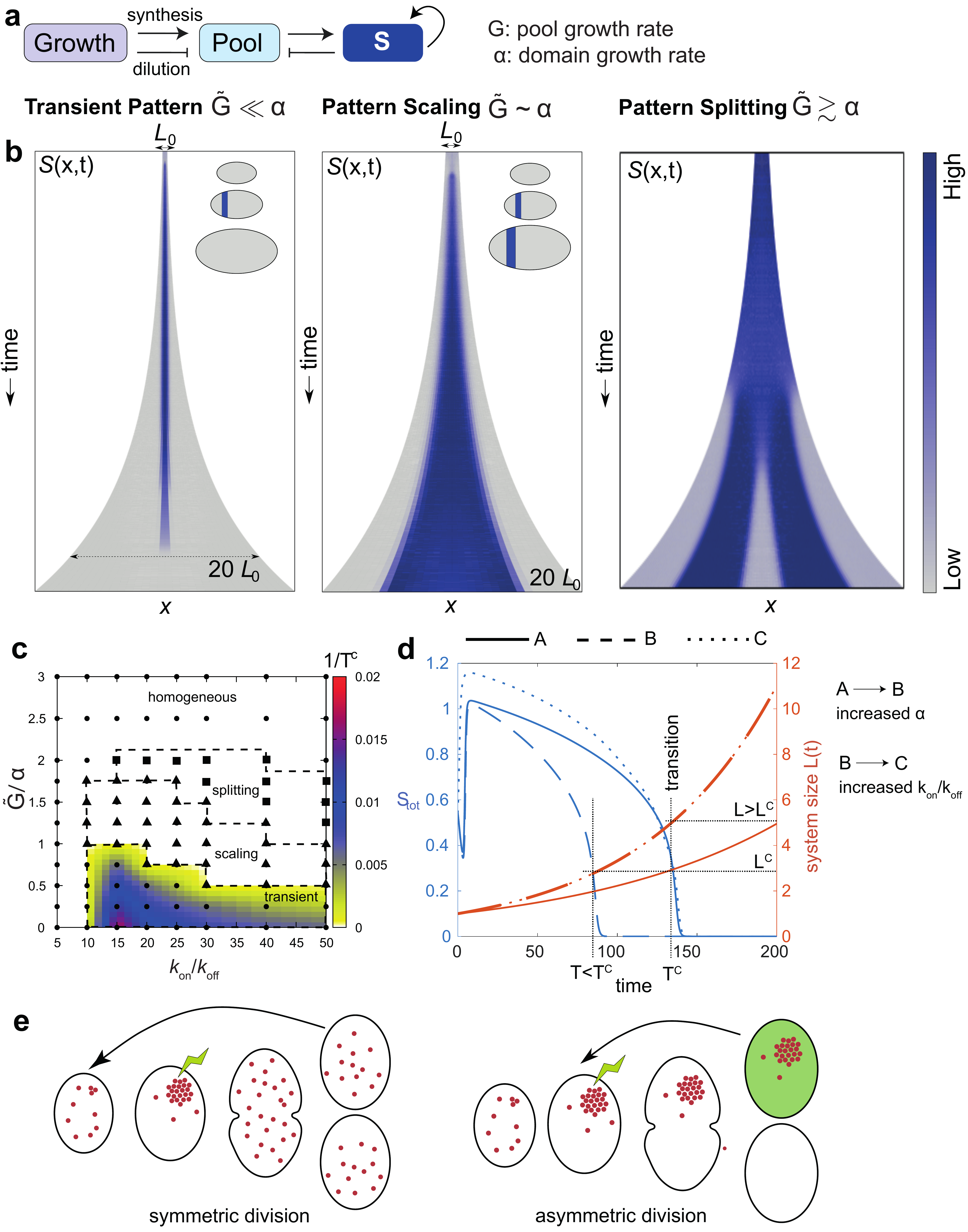}
\caption{{Pattern scaling, splitting and transient pattern formation in growing domains. (a) Feedback motif for an activator-substrate system coupled to a growing domain. (b) (Left) When subunits are produced at a rate slower than the rate of domain growth, growth-induced dilution leads to transient pattern establishment. (Middle) When the production of subunit pool occurs at a rate comparable to system size growth, the pattern formed grows in proportion to system size, exhibiting a dynamic scaling behaviour. This is different from sequential pattern formation as the polarity is preserved during growth. (Right) In the case of strong autocatalysis of $S$, the pattern spontaneously splits. (c) Phase diagram for pattern formation as functions of pool growth rate relative to the system, $\tilde{G}/\alpha$, and $k_\text{on}/k_\text{off}$. Colormap denotes the inverse of the pattern lifetime, $1/T^c$. (d) Time evolution of structure size $S_\text{tot}=\int_0^L S dx$ (blue) and system size $L$ (red) for the case of transient pattern formation. The lifetime of the pattern $T^c$, is coupled to the system size at transition to the homogeneous state, $L^c$. They can be tuned independently of each other, for example, by changing growth rate $\alpha$ (case B) where only the transition time changes, or by changing autocatalysis rate (case C) where $T^c$ remains the same but transition happens at a different system size. (e) Tunability of pattern lifetime can be utilised as a control mechanism for symmetric and asymmetric cell division (in terms of polarity protein content). When the pattern is transient (left) the dissolution of the structure will make the daughter cells symmetric in fate, containing the same amount of polarity proteins. If the pattern persists (right) then the division will lead to asymmetric fate inheritance. Parameter values: $D_P=1$, $D_S=0.005$, $\kappa_0=0.85$, $S_0^2=10$, $\alpha=0.01$, $L(0)=1$, and total pool density $\int(P+S)dx/L = 2$, with $G$ and $k_{\text{on}}/k_{\text{off}}$ variable.}
}
\label{fig:3}
\end{figure}

Pattern forming systems that align with activator-substrate or activator-inhibitor motifs are able to undergo sequential transitions in pattern concomitant with domain growth (Fig.~\ref{fig:2}-\ref{fig:2b}). As domain growth continues, patterns undergo further bifurcations to establish periodicity. While these motifs allow irreversible transitions in pattern, it may be desirable for systems to sense intermediate sizes, and for these transitions to be reversible. One can conceive of a timer-like set-up in a biphasic scheme: in the assembly stage, symmetry is broken at a critical size; and in the proceeding dissolution stage, patterns are lost at some larger size. To investigate the emergence of timer-like behavior we couple an activator-substrate system to a growing domain (Fig.~\ref{fig:3}a). %We find that a timer-like control of patterns is an emergent property of growing systems that couple pool growth to system size. By tuning the competition between domain growth and the production of subunit pool, one can induce transient pattern formation, pattern splitting or patterns size scaling with system size. 

%{\color{red} Can we not associate domain growth with sequential pattern formation? Some times it becomes confusing if we are talking about dynamic patterns in a time dependent domain size or a static pattern formation at different domain sizes (which arises in different stages of growth). Though the reviewers have not commented on this but I think it might be important to keep this in mind. -Deb
%Good point. I think the form we have is OK however. Shila, what do you think? - JCS}

%Figure illustrates various biological models of pattern formation in developing systems, including growing cells, tissues or organs  \cite{newman1979, miura2000}. In such cases, pattern formation and symmetry breaking occurs dynamically and the nature of the resulting polarity pattern depends on the process of growth \cite{kulesa1996,kondo1995,crampin1999}. Here we present few simple cases of symmetry breaking in activator-substrate reaction-diffusion models,    

%As before, we model the spatiotemporal dynamics of a regulatory structure $S$ in a system of size $L$, and coupled to a finite pool of building blocks. 
Specifically, we consider isotropic growth of the domain \cite{crampin1999} (all parts of the domain grow in a similar fashion) and the subunit pool density $P$ grows homogeneously with a constant rate. Due to domain growth, both the system size $L(t)$ and the total amount of building block pool, $N(t)=\int {(S+P)} dx$, are time-dependent. The growth of the system introduces local flow and dilution of both $S$ and $P$ \cite{crampin1999}. The coupled dynamics of $S$ and $P$ are given by (Fig.~\ref{fig:3}a):
\begin{equation}
\partial_t S + \frac{\dot{r}}{r}\left( x \partial_x S + S \right)=D_s \partial_x^2S + k_\text{on} P f(S) - k_\text{off} S\;,
\end{equation}
\begin{equation}
\partial_t P + \frac{\dot{r}}{r}\left( x \partial_x P + P \right) =D_p \partial_x^2 P - k_\text{on} P f(S) + k_\text{off} S + G\;,
\end{equation}
where $G$ is the growth rate of the subunit pool and the domain growth function $r(t)$ is defined as: $L(t) = L(0) r(t)$, where $L(0)$ is the initial system size. We write the assembly rate function as $f(S)=\kappa_0 + S^n/(S^n + S_0^n)$, where $\kappa_0$ defines the size-independent rate of assembly of $S$. Motivated by exponentially growing cells and tissues, we specifically consider the case of exponential growth in system size, such that $r(t) = e^{\alpha t}$ with $\alpha$ the growth rate. Since the macromolecular composition of cells scales with the cell size, we assume that the total amount of building blocks, $\int P dx$, grows at a rate proportional to system size $L$, resulting in a constant rate of growth $G$. While the formation of patterns occurs beyond a critical system size $L>L^*$, the stability of the pattern depends on the interplay between the rates of growth-induced dilution, synthesis of the subunit pool $P$, and autocatalysis of $S$.

\subsection*{Case 1: Transient polarity pattern due to growth-induced dilution}

When the subunit pool grows at a rate much slower than the rate of system growth $\tilde{G}\ll \alpha$ (where $\tilde{G}=GL(0)$), the structure is formed transiently and dissolves after a critical time $T^c$. The initial slow growth of the system size allows the formation of pattern beyond a critical size $L^*$. However, as $dL/dt$ increases rapidly (due to exponential growth) and becomes much faster than the rate of pool synthesis, the subunit density starts decreasing. Below a critical density of subunits, the pattern dissolves and the system reaches a homogeneous state (Fig.\,\ref{fig:3}b, left). Transient structure formation has been observed in slime mold \cite{postma2003} and during mammalian development \cite{erzurumlu1990}, and modelled using stochastic RD systems \cite{hecht2010}. Here we argue that system growth can also induce such transient polarity formation. 

The patterned state makes a transition to a homogeneous state at a time $T^c$ when the system size reaches $L^c$. The critical time for the transition to the homogeneous state, $T^c$, is determined by the parameters of the feedback motif ($k_\text{on}/k_\text{off}$, $\kappa_0$) and the growth rates $\alpha$ and $\tilde{G}$ (Fig.\,\ref{fig:3}c). The lifetime of the pattern, $T^c$, and the system size at transition to the homogeneous state, $L^c$, can be tuned independently of each other by modulating $\kappa_0$, $\alpha$ and $\tilde{G}$ (Fig.\,\ref{fig:3}d). Controlling the lifetime of polarity patterns is essential for regulating developmental transitions, and further experimental studies are essential to uncover such control mechanisms.  

\subsection*{Case 2: Pattern scaling due to proportional growth of system size and the subunit pool}

When the subunit pool grows at a rate comparable to the rate of growth of system size, $\tilde{G}\sim \alpha$, the subunits can reach a homeostatic density in time. As a result, the patterns formed during growth do not dissolve. Strong autocatalytic growth prevents delocalization of the early pattern and prevents the possibility of period doubling as seen in Schnakenberg kinetics and Gierer and Meinhardt model \cite{crampin1999}. This leads to a dynamic pattern scaling behaviour where the size of the pattern scales with the size of the system (Fig.\,\ref{fig:3}b, middle). This mechanism of scaling is notably different from the morphogen gradient scaling \cite{benzvi2011}. Here, pattern scaling is a consequence of system growth where the polarity pattern that does not change qualitatively with system size. 

\subsection*{Case 3: Pattern splitting}
When the rate of autocatalysis is sufficiently high, $\kappa_0\neq 0$, and the subunit pool and system size grow at similar rates, $\tilde{G}\gtrsim \alpha$, dynamic pattern splitting can emerge in the context of growth (Fig.\,\ref{fig:3}b, right). Here, slower positive feedback weakens the long-range inhibition arising as the consequence of pool depletion, allowing new peaks to emerge, in contrast to case 2. Adaptive benefits of pattern splitting may be multiple, for example allowing for the emergence of sequential patterning, and in maintaining relative stasis in local patterns upon domain extension. 

\section{Using growth as a timer: transient symmetry breaking at intermediate size}

%Systems that align with case 1 of our minimal model are able to undergo qualitative transitions in pattern concomitant with domain growth. As this domain growth continues, they undergo further bifurcations to establish periodic patterns. While these motifs allow irreversible transitions in pattern, which appear to be utilised in abundance across life, it may be desirable for systems to sense intermediate sizes, and for these transitions to be reversible. One can conceive of a timer-like set-up in a biphasic scheme: in the assembly stage, symmetry is broken at a critical size; and in the proceeding dissolution stage, patterns are lost at some larger size. 

Analysis of our minimal model shows that a timer-like control of pattern is an emergent property of symmetry-breaking systems that couple pool size to system volume. 
%Besides the simplicity of this motif (Fig.\,\ref{fig:3}a), upon a thorough literature search, we find that such a mechanism is seemingly less abundant than systems analogous to sequential pattern formation. 
We speculate that transient symmetry breaking may serve as a control strategy to mediate shifts between symmetric and asymmetric cell division in stem cell homeostasis. In particular, the biphasic nature of transient symmetry breaking scheme can help rationalise the equal distribution of determinants in the face of unequal nature of cell division. Suppose this system underlies the establishment of cell polarity required for asymmetric division. A critical cell size would allow polarity to be established, preventing precocious cell division. After the cell-size timer has elapsed, and the cell enters the dissolution phase, polarity proteins will return to the fast-mixing pool (Fig.\,\ref{fig:3}e, left). If cell polarity can have some effect on differential daughter cell fate, such a scheme would dissolve any bias prior to division.  

The significance of a mechanism like this can be understood in cases where cell lineages undergo switches between symmetric and asymmetric cell division. Cases 1 and 2 of our model are identical besides the relative rates of pool synthesis, system size growth and autocatalysis. Hence tuning the coupling between pool production rate and system size can regulate transitions between reversible polarisation (case 1) and irreversible polarisation (case 2), wherein polarity is maintained through a cell division event, maintaining a bias in determinants and seeding differential daughter-cell fate (Fig.~\ref{fig:3}e, right). In situations where such switches between symmetric and asymmetric division are dynamic, regulating the extent of growth-induced dilution to effect switches between transient versus irreversible polarisation may be an optimal control strategy. 

%Research over the past decade has thrown into question the paradigm of asymmetric cell division as the mechanism underpinning stem cell maintenance \cite{clayton2007,snippert2010, lopez2010}. Lineage tracing studies and quantitative models have demonstrated that, across several mammalian homeostatic tissues, many cells have the \textit{capacity} to self-renew and give rise to differentiated lineages — the defining characteristics of a stem cell — but only a small subset actually \textit{realise} this capacity. Instead of asymmetry in fate at the single cell scale, tissue homeostasis is achieved by population asymmetry, suggesting that while the balance of self-renewing and differentiated cells is maintained across the tissue, the fates of individual cells are unpredictable. In a process of neutral competition, some will realise their capacity to replenish differentiated lineages, whereas others lose out by chance. 

%Population asymmetry appears to be abundant in stem cells that show mixed modes of symmetric and asymmetric division \cite{klein2011a}, for example in haematopoeitic precursor cells \cite{wu2007} and in muscle satellite stem cells \cite{kuang2007}, although this mixed mode is not theoretically necessary \cite{simons2011}.  While the shift division modes could be realised post-mitosis, whereby daughter cells negotiate the fates of one another (e.g. seen in {\it Drosophila} intestinal stem cells \cite{de2012}), an equally plausible control strategy could decide alternate division modes prior to mitosis. 

Given the switch between reversible and irreversible symmetry breaking in our minimal model is governed by a single parameter change, it is tempting to speculate that such a mechanism may be responsible for the mixed modes of stem cell proliferation, which in many systems are seemingly stochastic. Under such a scenario, the switch between these division modes may be noisy at the level of individual cell decisions, given the requirement of being poised near the transition point. However, given fate decisions and patterns of cell division are known to be influenced by signals emanating from stem cells or their progeny in many systems, such a model would confer plasticity and robustness in stem cell homeostasis at the level of the population. We stress that this mechanism remains a theoretical prediction and are intrigued as to whether such a control strategy is indeed utilised in nature. 

%\textcolor{red}{JCS: Have amended this. What do you think?}

\section{Overcoming size constraints: scaling patterns in growing systems}

Not all biological systems that display symmetry breaking also show size-dependent pattern formation. Indeed, it may be adaptive for systems to canalise their patterning mode irrespective of size. This is a feature of case 2 of our minimal model (Fig.~\ref{fig:3}b, middle), which features commensurate growth of pool size and system size, leading to scale invariance upon domain growth: if the system breaks symmetry to form a single structure at a smaller size, upon isotropic growth, the system maintains a single structure which grows in proportion to the domain as a whole. This motif utilises the symmetry breaking capacity of reaction-diffusion systems but subverts the feature of intrinsic wavelengths characteristic of traditional Turing circuits. In this section, we delineate two potential modes of scale-invariant symmetry breaking systems — one with history dependence, and one without — and argue that such systems display adaptive features in certain contexts. 

%Scale-invariance in morphogen gradient establishment is both abundant in nature ~\cite{lesne2012} and has become a field of active theoretical research in recent years, with for example the proposal of an integral feedback loop (expander-repressor motif) in regulating gradient shape in line with domain size ~\cite{ben2010, ben2011}. However, scale-invariance in self-organising systems has been less extensively explored. 

\subsection{Autocatalysis as a mechanism to preserve patterns in the face of growth}

Patterning in developing systems is almost invariably proceeded by growth, which is often proportional to the initial pattern. Traditional morphogen gradient hypotheses~\cite{wolpert1969} have implicitly assumed that the tissue is initially patterned when it is small and subsequently undergoes growth, facilitating proportional extension of the pattern. This two-phase model of patterning, where cell fates are assigned during an initial patterning phase, face the challenge of noise: while growth can lock in the lower positional error entailed by patterning in small fields of cells, this error cannot be reduced through growth. Accordingly, small errors in boundary position can be amplified upon growth, demanding the read-out of positional information at early stages is exquisitely tuned. If however fates are assigned in a self-organised manner, as in symmetry-breaking the systems we overview in our minimal model, this hard limit on noise in boundary positioning can be surpassed. Provided the symmetry breaking system scales with domain size, absolute noise in boundary position if anything reduces with growth; self-organised systems such as these continually refine boundaries throughout growth, rather than amplifying noise in initial specification. 

Case 2 of our minimal model allows for pattern scaling via proportional growth of pool and domain size, and autocatalysis, which in effect instills history-dependence in pattern formation, thus helping to preserve proportions. Hence the pattern generated at small domain sizes is preserved upon elongation. Given the diffusion length scale shortens with respect to relative domain size upon growth, boundaries sharpen over time. In the context of a developing field of cells, this autocatalysis could represent positive feedback in master transcription factors or indeed epigenetic changes, which allow cells and their progeny to remember past states. Hence such a model may help provide alternative mechanisms for scaling of patterns with growth, whereby initial stages establish the crude pattern (e.g. number and position of structures), which is in turn refined over time. The hallmark of actively scaling processes such as these is the reduction of noise in boundary positioning, i.e. violating the data-processing inequality~\cite{kinney2014}.

%While autocatalysis-mediated pattern scaling is more robust to boundary position noise, two-phase models show adaptive benefits when growth is differential and hence contributes to patterning itself. Indeed, this has been seen in the developing neural tube, where proliferation is anisotropic, and different cell identities display differential dynamics of differentiation~\cite{kicheva2014}. 

\subsection{Expander-coupled systems can scale patterns to domain size irrespective of history}

While symmetry-breaking schemes incorporating autocatalysis show benefits of maintaining patterns with growth, certain systems may require patterning to be scale-invariant without the requirement of time-dependence. This is exemplified in regenerating systems, which are able to regrow organs or entire organisms in the correct proportions, in spite of drastically different starting sizes. Recent theoretical work has advanced understandings of how scale-invariant symmetry breaking could operate. Werner et al.~\cite{werner2015} proposed that a third component is required, analogous to expanders in morphogen gradient scaling, which dynamically modulates patterning wavelength as a function of system size by tuning levels of pattern forming molecules. This model demonstrated time-independent scaling across several orders of magnitude differences in domain size. While the model is based on a traditional activator-inhibitor model, the scheme is generalisable to other modes of expander-mediated modulation and other symmetry breaking motifs such as substrate depletion. 

%This model bears some concordance with the regeneration of planarians. This taxon has a remarkable capacity for regeneration, with small dissected fragments of tissue showing the capacity to remake the entire organism. Key to this is re-patterning growing tissues to reestablish organismal polarity. The anterior-posterior axis of planarians is dictated by graded Wnt signalling, highest toward the head~\cite{sureda2016}. Stückemann et al.~\cite{stuckemann2017} found that the gradient of Wnt signalling scales with body size during regeneration, and in knock-down treatments that dwarf body lengths. Thus unlike patterning paradigms in the development of higher organisms, where patterning is thought to be limited to a narrow time-window, planarians appear to constitutively self-organise axis information. Such work contrasts with previous Turing-based models of regeneration (in \textit{Hydra}) that posit a third factor — a “source density” — that retains a memory of pattern upon dissection. This factor was proposed to be highest at the head, but had much slower kinetics, allowing for graded information to be carried through to dissected organismal fragments. Double-tailed planarians are equally capable of displaying graded Wnt signalling in line with their morphology, arguing against the role of specialised regions (e.g. the head) in guiding fate. Thus axis regeneration in planarians is a bone-fide example of self-organised patterning. 

\section{Discussion}

In this Perspective, we presented a minimal model for symmetry breaking to serve as a unifying framework to understand pattern formation in the context of timing and growth. We argue that systems that display positive feedback in activator recruitment, drawn from a limiting pool, can yield spontaneous symmetry breaking. This basic scheme is mathematically akin to other RD mechanisms including activator-inhibitor or substrate-depletion motifs, all relying on a common logic of local activation and long-range inhibition. Thus the insights gleaned from the phenomenological behaviour of this system is applicable to diverse systems. 

Across the cases of the minimal model we consider, we observe a hard size limit on pattern formation: below a critical size, diffusive dispersion overwhelms the capacity to break symmetry. Given developing systems across scales typically display patterning and growth occurring in unison, if this critical size is within biologically meaningful length scales, such behaviour can elicit qualitative changes in patterning: growth above a critical size leads to a bifurcation, whereby the system transitions from unpolarised to polarised. Viewing growth as a control parameter of the system that increases system size at a predictable rate, developmental systems can utilise this bifurcation to enact developmental transitions at the right place and time. 
%In this view, taking growth as a given, developmental systems can measure time and thus orchestrate changes in form in an ordered manner. 
As a generic by-product of symmetry breaking systems, we predict that this time-keeping mechanism may be more abundant than anticipated. We note that this feature is the most generic among RD models of pattern formation, and among the different cases of our minimal model: the diffusion length-scale sets a physical limit on pattern formation. 

Beyond this first bifurcation, our mass-conserved RD model predicts different dynamic behaviors depending on the regulatory motifs. These include sequential pattern formation, transient pattern formation, pattern scaling, and pattern splitting in growing systems. In line with the well-established literature on domain size in Turing patterns~\cite{koch1994}, our model predicts \textbf{clock-like} sequential pattern formation (Fig.~\ref{fig:2}-\ref{fig:2b}): as the system grows larger, given patterning wavelength is intrinsic to the system, the domain can accommodate multiple structures. An important dynamical consequence of this is \textit{temporal ordering}: growth elicits consecutive bifurcations, resulting in sequential patterning, shown to be instrumental in neuronal cell (bi)polarity~\cite{menchon2011}, and joint patterning in digits~\cite{cornwallscoones2020}. Alternatively, growth-induced pool dilution can drive systems back towards an unpolarised state, allowing for \textit{transient pattern formation} at intermediate size (Fig.~\ref{fig:3}). Thus an alteration in growth regulation can yield {\bf timer-like} dynamics, which we hypothesise may be important in orchestrating switches between asymmetric and symmetric stem cell division modes. A qualitatively different behaviour upon continued growth is \textit{scale-invariance}, whereby the proportions of the pattern are maintained upon domain elongation. Scale-invariant systems show \textbf{switch-like} dynamics, becoming time-independent after the first bifurcation. We argue that such behaviour could allow patterned tissues to maintain proportions upon proliferation, where self-organisation continually refines boundary position instead of stretching noise in initial specification. 
%We delineate two possible modes of scale-invariance in symmetry breaking: one utilising autocatalysis, via a memory-like effect, suited for developing tissues; and another requiring a third component, an expander, to dictate pattern in a time-independent manner, suited for regenerating tissues. 

%Thus changes in regulatory motifs can yield qualitatively different responses upon continued growth. 

Our reaction-diffusion framework for understanding developmental time in terms of size-dependent symmetry breaking is generalisable beyond the systems that couple increases in size to developmental transitions via biochemical circuits. Firstly, decreases in system size can also be utilised by developmental systems to temporal transitions. For example, the transition from asymmetric to symmetric division in the P-lineage of {\it C. elegans} can be understood in terms of sequential reductions in cell volume pushing the system over the critical cell size threshold for polarisation. Secondly, the organising principle of Turing-like pattern formation — local-activation and long-range inhibition — extends beyond systems based solely on chemical cross-talk~\cite{hiscock2015}: pattern formation can emerge from cell-cell interactions or mechanical instabilities~\cite{mark2010,banerjee2015,ravasio2015}. While we restricted our focus in this paper to biochemical systems, future work should attempt to unify these results with mechanically driven size-dependent symmetry breaking. Indeed, we may see strong parallels in how nature utilises chemical or mechanical instabilities to regulate the timing of developmental transitions. We hope that our proposed strategies for time-keeping in natural living systems can also provide inspirations for engineering of synthetic circuits with tunable dynamics. 

\vspace{6pt} 

%%%%%%%%%%%%%%%%%%%%%%%%%%%%%%%%%%%%%%%%%%
%% optional
%\supplementary{The following are available online at \linksupplementary{s1}, Figure S1: title, Table S1: title, Video S1: title.}


%%%%%%%%%%%%%%%%%%%%%%%%%%%%%%%%%%%%%%%%%%
%\authorcontributions{Conceptualization, J.C.S., D.B. and S.B.; methodology, D.B.; validation, J.C.S., D.B. and S.B.; formal analysis, D.B.; investigation, J.C.S., D.B. and S.B.; resources, S.B.; data curation, J.C.S., and D.B..; writing--original draft preparation, J.C.S., D.B. and S.B.; writing--review and editing, J.C.S., D.B. and S.B.; visualization, J.C.S., D.B. and S.B.; supervision, S.B.; project administration, S.B.; funding acquisition, S.B. All authors have read and agreed to the published version of the manuscript.}
\pagebreak
%%%%%%%%%%%%%%%%%%%%%%%%%%%%%%%%%%%%%%%%%%
\acknowledgments{The authors thank Andrew Goryachev for many useful comments. SB acknowledges funding from Royal Society University Research Fellowship URF/R1/180187, and Human Frontiers Science Program (HFSP) grant number RGY0073/2018.}

%

%%%%%%%%%%%%%%%%%%%%%%%%%%%%%%%%%%%%%%%%%%
\end{document}